\documentclass[11pt,tightenlines,eqsecnum,floats,aps,amsmath,amsart,superscriptaddress,amssymb,nofootinbib,prd,shownopAshtekar:2007em,floatfix]{revtex4}
\usepackage{graphicx, wrapfig}
\usepackage{amssymb}
\usepackage[usenames, dvipsnames]{color}
\usepackage{mathrsfs}
\usepackage{subfigure}
\usepackage{graphicx}
\usepackage{verbatim}
\usepackage{cancel}
\usepackage[normalem]{ulem}
\usepackage[hidelinks]{hyperref}
\hypersetup{
  colorlinks   = true, 
  urlcolor     = blue, 
  linkcolor    = blue, 
  citecolor   = blue 
}
\def\dd{\textrm{d}}

\def\u{\mathfrak{A}}
\def\dd{\textrm{d}}

\def\Q{\hat{\mathcal Q}}

\usepackage{enumerate}

\newcommand{\ig}{\includegraphics}
\newcommand{\be}{\nopagebreak[3]\begin{equation}}
\newcommand{\ee}{\end{equation}}
\newcommand{\bfig}{\nopagebreak[3]\begin{figure}}
\newcommand{\efig}{\end{figure}}
\newcommand{\bea}{\nopagebreak[3]\begin{eqnarray}}
\newcommand{\eea}{\end{eqnarray}}

\newcommand{\bmult}{\nopagebreak[3]\begin{multline}}
\newcommand{\emult}{\end{multline}}

\begin{document}

\title{Primordial power spectrum from the Dapor-Liegener model of loop quantum cosmology}
\author{Ivan Agullo}\email{agullo@lsu.edu}
\affiliation{Department of Physics and Astronomy, Louisiana State University, Baton Rouge, LA 70803, U.S.A.}

\begin{abstract}
 We compute the predictions for the power spectrum of scalar perturbations from a recent new proposal for the effective Hamiltonian of loop quantum cosmology. The model provides an attractive picture of the early cosmos, in which our classical Friedmann-Lema\^itre-Robertson-Walker universe emerges from a quantum phase where the spacetime curvature remains constant and of Planckian size.  We compare the predictions for the cosmic microwave background with previous results obtained within loop quantum cosmology, and discuss the differences and similarities. The analysis provides an example of the way  differences between quantization schemes can be translated to physical observables. 
 
\end{abstract}


\maketitle

\section{Introduction}

The  standard model of cosmology provides a satisfactory description of our universe, with predictions that agree with observations at the percent level. But this model is incomplete from a fundamental viewpoint, since it  rests  on Einstein's general theory of relativity. In the extreme regime found in the Planck era, quantum gravity effects cannot be neglected. In the absence of a complete theory of quantum gravity,  insights about the Planck era have been obtained by using approximations and extrapolations of  ideas available at the present time. Examples are the no-boundary proposal \cite{Hartle:1983ai}, string cosmology \cite{Lidsey:1999mc}, and loop quantum cosmology \cite{Ashtekar:2011ni}, just to mention a few. But the lack of a complete theory unavoidably makes mathematical ambiguities to appear,  that require choices to be made in order to make physical predictions. 
Progress in physics has always been driven by an interplay between theory and observations, and cosmology provides a promising scenario to confront theoretical ambiguities in quantum gravity with real data.

This paper focuses on the approach provided by loop quantum cosmology (LQC); see \cite{Bojowald:2008zzb, Ashtekar:2011ni,MenaMarugan:2011va,Banerjee:2011qu,Agullo:2013dla,Calcagni:2012vw,Barrau:2013ula,Ashtekar:2015dja,Agullo:2016tjh,Grain:2016jlq,ElizagaNavascues:2016vqw,Wilson-Ewing:2016yan,Barrau:2016nwy} for reviews. 
In this framework,  quantum effects make the big bang singularity to be replaced by a quantum bounce \cite{Ashtekar:2006rx,Ashtekar:2006uz,Ashtekar:2006wn}. Before and after this bounce, the spacetime curvature  decreases rapidly in cosmic time, quantum effects of gravity weaken, and general relativity quickly becomes  an excellent approximation. In this sense, the bounce provides a `quantum bridge' between  two classical universes, one contracting and one expanding. These results have been derived for homogeneous and isotropic models, with and without spatial curvature \cite{Ashtekar:2006rx,Ashtekar:2006uz,Ashtekar:2006wn,Ashtekar:2006es,Szulc:2006ep} and  cosmological constant \cite{Pawlowski:2011zf}, anisotropic models \cite{Ashtekar:2009um,Ashtekar:2009vc,WilsonEwing:2010rh}, as well as in inhomogeneous Gowdy models \cite{Garay:2010sk,MartinBenito:2010bh}.

The formulation of loop quantum cosmology is not free of quantization ambiguities. The original approach in \cite{Ashtekar:2006rx,Ashtekar:2006uz,Ashtekar:2006wn} was to use physical arguments to discriminate among  possibilities, making sure that the final picture is physically viable.  The field has matured significantly since then, and a robust physical  picture with predictions for the cosmic microwave background temperature anisotropies is now available \cite{Agullo:2012sh,Agullo:2013ai,Agullo:2015tca,MenaMarugan:2011me,deBlas:2016puz,Ashtekar:2016wpi,Bonga:2015kaa,Agullo:2016hap,Barrau:2013ula,Zhu:2017onp,Wilson-Ewing:2016yan}. Cosmological observations have been used to reduce the space of free parameters of the theory, and additional predictions involving tensor perturbations and non-Gaussianity \cite{Agullo:2017eyh,Agullo:2015aba} have been made with the goal of testing the basic ideas.

With the experience accumulated so far in LQC, it is natural to look back at the fundaments and  study whether there are other mathematical possibilities that would make the theory still viable, and perhaps closer to  full loop quantum gravity. An interesting recent  proposal in this direction has been made by Dapor and Liegener  in \cite{Dapor:2017rwv,Dapor:2017gdk}, following previous work by Yang, Ding and Ma \cite{Yang:2009fp} (see also \cite{Alesci:2013xd,Alesci:2016xqa,Alesci:2017kzc} for a different, interesting proposal).  By applying Thiemann's proposal to  regularize the Hamiltonian constraint of loop quantum gravity, 
Dapor and Liegener have derived an effective Hamiltonian for LQC, by computing the expectation value on suitably defined complexified coherent states that describe homogeneous and isotropic FLRW quantum geometries. This proposal has been formulated in more solid grounds  in \cite{Assanioussi:2018hee}, and  Li, Singh, and Wang \cite{Li:2018opr} have provided a detailed and careful analysis of its effective spacetime dynamics. For concreteness, we will refer to this new proposal for LQC dynamics as the Dapor--Liegener  dynamics, to differentiate it from the standard approach to LQC. 

The goal of this paper is to investigate the consequences of the Dapor--Liegener proposal for the cosmic microwave background (CMB). More precisely, we compute the primordial spectrum of scalar perturbations, and compare it with the one obtained in standard LQC. We will address the following question: (i) Is the new proposal compatible with existing observational constraints? (ii) Does  it produce any effect in the CMB  that will allow us to differentiate it from other models? (iii) Does this new proposal sheds any new light on the choice of the initial state for cosmological perturbations? 

The viewpoint  in this paper is purely phenomenological. In other words, rather than investigate the mathematical foundations of the model, we will focus on its observational consequences. The sprit here is to bring different mathematical possibilities in loop quantum cosmology to the realm of observations. 

Section 2 contains a pedagogical overview of the effective spacetime dynamics predicted by the Dapor--Liegener model. We discuss in section 3 the new elements that this models provides on the discussion of initial quantum state for cosmological perturbations, compute the primordial power spectrum of scalar modes, and compare the result with other models in the literature. Hence, section 3 contain the main analysis of this paper. Section 4 contains a summary and a discussion of the main results. 

Along this paper we use $c=1$, and Planck units for mass ($m_{P\ell}\equiv\sqrt{\hbar/G}$) and time ($t_{P\ell}\equiv\sqrt{\hbar\, G}$),

\section{Effective background dynamics}

This section provides a brief overview of the effective dynamics derived from the Dapor--Liegener  proposal for LQC. Additional details can be found in \cite{Assanioussi:2018hee,Li:2018opr}.

Consider the classical phase space  of spatially-flat FLRW gravitational fields, sourced by  a scalar field $\phi$ with potential $V(\phi)$. This phase space is four-dimensional. In general relativity, it is common to use the scale factor $a$, the scalar field $\phi$, and their canonically conjugate variables $\pi_a$ and $p_{\phi}$ as coordinates: $(a,\pi_a;\phi,p_{\phi})$.  In LQC, however, it is more convenient to use the  variables $v$ and $h$ for the gravitational sector, defined as\footnote{Because we are working here with homogenous fields in a spatially non-compact spacetime, the spatial integrals involved in the definition of the Hamiltonian and the symplectic form diverge.  But this is a spurious mathematical divergence. It can be eliminated by restricting the integrals to some finite, although arbitrarily large cubical coordinate volume $\mathcal{V}_0$. Physical observables do not depend on $\mathcal{V}_0$, and the limit $\mathcal{V}_0\to \infty$ can be taken at the end of the calculation.}  $v\equiv \frac{\mathcal{V}_0}{2\pi \, G}\, a^3$, and $h\equiv-\frac{4\pi G}{3a^2}\pi_a$. The non-zero Poisson brackets between these variables are
\be \{h,v\}=2\  ,\ \ \ \  \{\phi,p_{\phi}\}=1\, .\ee
When evaluated on solutions of the classical Friedmann equations, $h$ equals the Hubble rate $H\equiv \dot a/a$, but this is no longer true in loop quantum cosmology.\footnote{The relation between $v,h$ and the  variables $c$ and $p$ that are also commonly used in LQC is $v= \frac{p^{3/2}}{2\pi \, G}$, $h=c/(p^{1/2} \gamma)$, where $\gamma$ is the Barbero-Immirzi parameter.} 

The Dapor-Liegener  effective Hamiltonian constraint  is \cite{Dapor:2017rwv,Assanioussi:2018hee}
\be \label{Heff} \mathcal{H}_{\rm eff}=\mathcal{H}_{\rm M}-\frac{3}{4}\, v\,\frac{\sin^2(\ell_0\, h)}{\ell_0^2}\, \Big[1-(1+\gamma^2)\, \sin^2(\ell_0\, h)\Big]\approx 0\, , \ee
where $\mathcal{H}_{\rm M}$ is the matter Hamiltonian, given by $\mathcal{H}_{\rm M}=2\pi G\, v\, \rho$, with $\rho$ the matter energy density $\rho=\frac{1}{8\pi^2G^2\, v^2} \, p_{\phi}^2+ V(\phi)$. 
The Hamiltonian (\ref{Heff}) depends on two constants, the Barbero-Immirzi parameter $\gamma$, and  $\ell_0=\gamma\, \sqrt{\Delta_0}$, where $\Delta_0$ is the `area gap' of  loop quantum gravity ($\ell_{Pl}$ is the Planck length). The most common value of $\gamma$ used in the literature is $\gamma=0.237$, and it is motivated by black hole entropy counting \cite{Meissner:2004ju,Agullo:2008eg,Agullo:2010zz,Agullo:2008yv}. The value of the area gap in loop quantum gravity is  $\Delta_0=4\pi\gamma\sqrt{3}\, \ell_{Pl}^2$. 

 The classical Friedmann-Lema\^itre Hamiltonian is recovered from (\ref{Heff}) by sending the Planck length $\ell_{Pl}$ to zero, or equivalently  $\ell_0 \to 0$. On the other hand, the standard effective Hamiltonian used in LQC is  obtained by simply removing the factor between square brackets in (\ref{Heff}).%

From (\ref{Heff}), one obtains the equations of motion in cosmic time via Hamilton's equations:
\bea \label{veq} 
\dot v&=&\{v,\mathcal{H}_{\rm eff}\}=\frac{3}{2}\, v\, \frac{\sin(2 \, \ell_0\, h)}{\ell_0}\, [1-(1+\gamma^2)\, 2\, \sin^2(\ell_0\, h)] \, , \nonumber \\
\dot h&=&\{h,\mathcal{H}_{\rm eff}\}=-\frac{1}{\pi G\, v^2} p_{\phi}^2 -\frac{2}{v}\, \,\mathcal{H}_{\rm eff}\, ,\nonumber \\
\dot \phi&=&\{\phi,\mathcal{H}_{\rm eff}\}=\frac{1}{2\pi G\, v}\, p_{\phi}\, , \nonumber \\
 \dot p_{\phi}&=&\{p_{\phi},\mathcal{H}_{\rm eff}\}=2\pi G\, v\, \{p_{\phi},V(\phi)\}\, . \eea

One can obtain a qualitative understanding of  the solutions to these equations without  solving them, as follows.  First, from the Hamiltonian constraint (\ref{Heff}), we have
\be \label{rhoh} \rho=\frac{3}{8\pi G}\,  \frac{\sin^2(\ell_0\, h)}{\ell_0^2}\, [1-(1+\gamma^2)\,  \sin^2(\ell_0\, h)] \, .\ee
We can see from here that $\rho$ is bounded from above. The maximum value of the energy density is $\rho_{\rm sup}=\frac{3}{32\pi G\ell_0^2(1+\gamma^2)}=0.097$ (in Planck units). Notice that this is a factor $\frac{1}{4(1+\gamma^2)}\approx 0.24$ {\em smaller that in standard LQC}. 

On the other hand, from the first equation in (\ref{veq}) we obtain an expression for the Hubble rate
\be \label{Hh} H\equiv \frac{\dot v}{3v}=\frac{1}{2}\,  \frac{\sin(2\, \ell_0\, h)}{\ell_0}\, [1-(1+\gamma^2)\, 2\,  \sin^2(\ell_0\, h)] \, .\ee
From here  we can obtain that $H$ is also bounded, by $H^2_{\rm sup}=\frac{\left(3\gamma^2+\sqrt{8+16\gamma^2+9\gamma^4}\right)^2\left[4+3 \gamma^4+\gamma^2\left(8+\sqrt{8+16\gamma^2+9\gamma^4}\right)\right]}{512\ell_0^2(1+\gamma^2)^2}=(0.53\, t_{P\ell}^{-1})^2$.

\bfig \begin{center}
 \ig[width=0.68\textwidth]{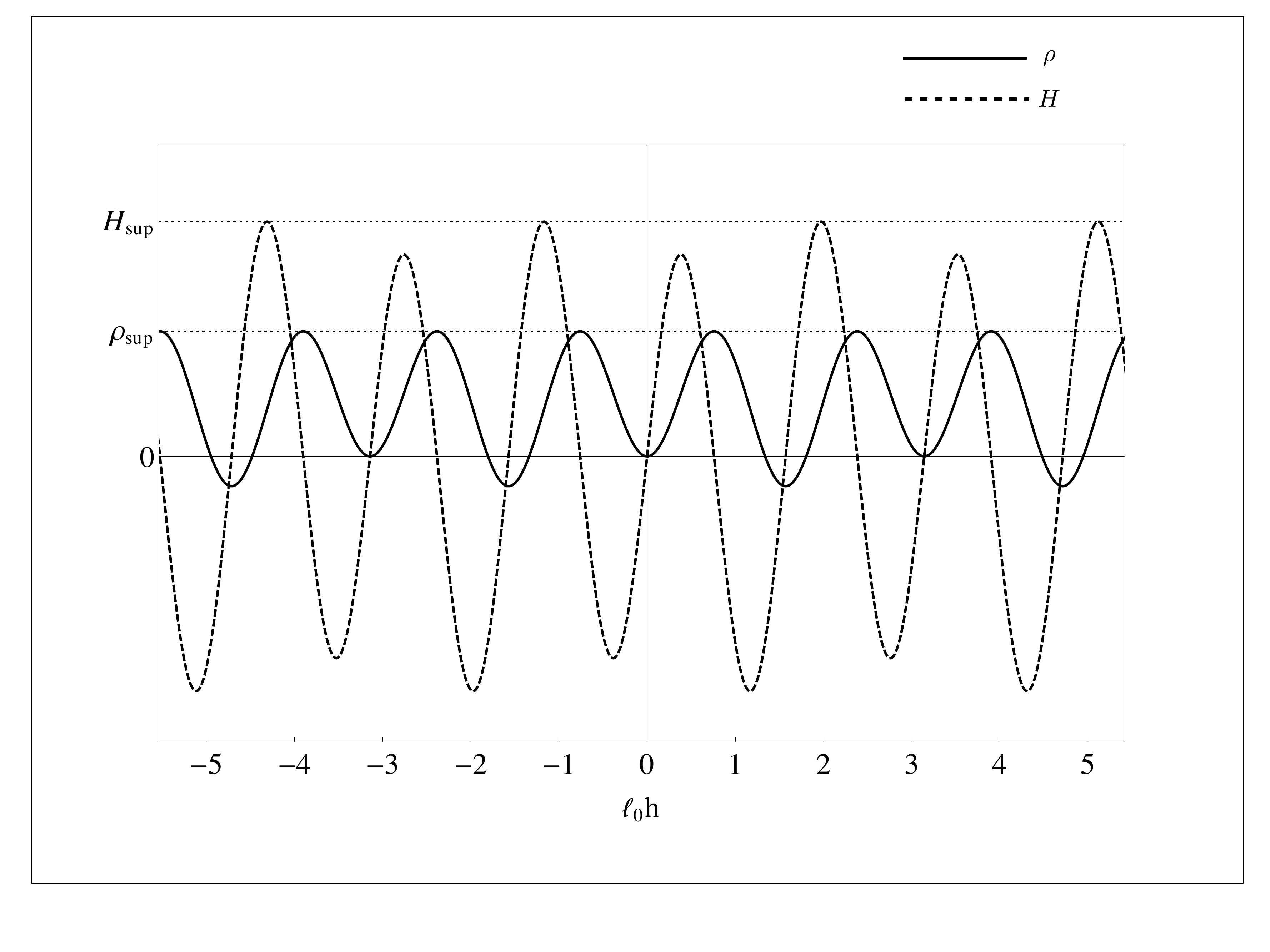}
\caption{Energy density $\rho$ (solid line) and Hubble rate $H$ (dashed line) versus $\ell_0\, h$. Both functions are periodic. The energy density, which is physically positive definite, takes negative values for some values of  $\ell_0\, h$. Therefore, the model will be physically consistent only if these values  of  $\ell_0 h$ are dynamically disconnected   from physically allowed regions.}
\label{H&rho}
  \ig[width=0.68\textwidth]{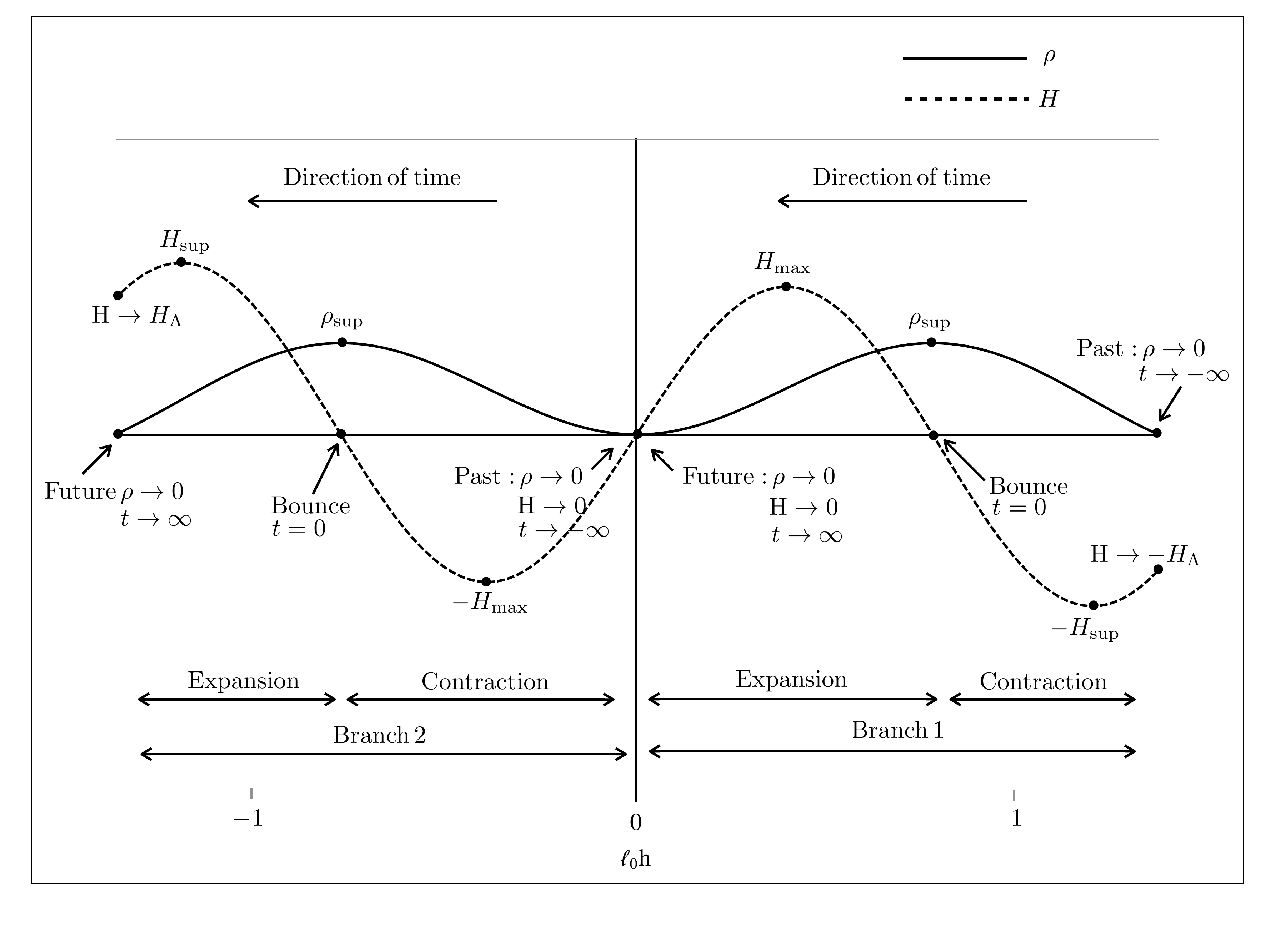}
\caption{Energy density $\rho$ (solid line) and Hubble rate $H$ (dashed line) versus $\ell_0\, h$ for one cycle with $\rho\geq 0$, that corresponds to $\ell_0 h\in[-\arcsin{(1/\sqrt{1+\gamma^2})},\arcsin{(1/\sqrt{1+\gamma^2})}]$. Cosmic time runs from right to left. In either region, $h>0$ (Branch 1) or $h<0$ (Branch 2), cosmic time runs from $-\infty$ to $+\infty$. Hence the two branches are {\em not} connected by dynamics. Branch 1 represents a universe that initially contracts as a FLRW-de Sitter spacetime with constant $H=-H_{\Lambda}$, bounces, and then expands towards vanishing $\rho$ and $H$. That is to say, quantum gravity produces an emergent cosmological constant {\em only} before the bounce. Branch 2 represents the time reversal solution. 
Note that this figure plots only $H(\ell_0 h)$ and $\rho(\ell_0 h)$, and  it does not contain information about the  amount of cosmic time elapsed between different points in the plot, i.e.\, there is no information about $h(t)$. Such information is contained in Figure \ref{Hvst}, for a particular solution to the equations (\ref{veq}). As an example of how non-trivial the function $h(t)$ is, notice that the Hubble rate remains constant and equal to $H=-H_{\Lambda}$  in the contracting phase of Branch 1 until just a few Planck-times before the bounce! (see figure \ref{Hvst}).}
\label{H&rhozoom2}\end{center}
\efig

Figure \ref{H&rho} shows the matter energy density $\rho$ and Hubble rate $H$ versus $\ell_0\, h$ [equations (\ref{rhoh}) and (\ref{Hh})]. This figure shows two relevant features. On the one hand, both $\rho$ and $H$ are periodic functions.  On the other, there are values of $\ell_0\, h$ for which $\rho$ becomes negative. The physical energy density is definite positive; therefore, physical consistency requires that the values of $\ell_0\, h$  for which $\rho$ becomes negative are never reached by dynamics. As we now explain, this is in fact the case. 

In order to study the evolution in more detail, figure \ref{H&rhozoom2} shows just a single cycle of $\rho$ and Hubble rate $H$. First of all, notice that the second equation in (\ref{veq}) shows that $h$  {\em decreases} monotonically with cosmic time ($\dot h\leq 0$). Hence, future evolution takes place in figure \ref{H&rhozoom2}  by moving from right to left. 

\begin{enumerate}
\item{\bf FLRW-de Sitter contraction}

 The rightmost part of figure \ref{H&rhozoom2} corresponds to cosmic time $t\to -\infty$, vanishing energy density  $\rho \to 0$, while the Hubble rate reaches a constant (negative) asymptotic value $-H_{\Lambda}\equiv-\frac{1}{\ell_0}\frac{\gamma}{1+\gamma^2}$. This corresponds to a contracting FLRW-de Sitter universe, with a scale factor that decreases exponentially fast, as $a(t)=e^{-H_{\Lambda}\, t}$. The curvature of the universe remains constant during this phase, and it is of Planck size, $R_{\Lambda}=12H_{\Lambda}^2=2.08 \, \ell_{P\ell}^{-2}$. Therefore, the  Universe acquires, in the asymptotic past, an effective cosmological constant  $\Lambda=3 \, H_{\Lambda}^2$ originated from quantum gravity effects, which remains active all the way back to $t\to -\infty$. This phase is very different from standard LQC, where the universe reaches a low curvature, classical phase in the past.
 
 The emergence of this  cosmological constant can be seen directly from equations (\ref{rhoh}) and (\ref{Hh}). In the limit $\ell_o h\to -\arcsin{(1/\sqrt{1+\gamma^2)}}$, that corresponds to $t\to -\infty$, equation (\ref{rhoh}) shows that $\rho\to 0$, while equation  (\ref{Hh}) reveals that $H\to -\frac{1}{\ell_0}\frac{\gamma}{1+\gamma^2}$. Note that the factor 2 multiplying $(1+\gamma^2)$ in (\ref{Hh}), which is absent in (\ref{rhoh}), is crucial in order to make $H$ a non-zero constant in the past.

 \item{\bf Quantum bounce}
 
While the universe contracts exponentially fast, energy density of the scalar field grows, and eventually  dominates over the effective cosmological constant. When this happens, $H$ does  not remain constant anymore, reaches a global minimum $-H_{\rm sup}$ and rapidly goes to zero. This is a cosmic bounce. Energy density reaches its maximum value $\rho_{\rm sup}$ at the time of the bounce. 

Numerical simulations show that the transition from de Sitter-like contraction to a bounce occurs {\em very close to the bounce},  about 6 Planck-times before it! In other words, the Hubble rate  reaches its past asymptotic value  very soon before the bounce. 
Notice that this information is not contained in figure \ref{H&rhozoom2}; one needs to integrate the equations (\ref{veq}) in order to understand how fast $h(t)$ changes with cosmic time in  different regions of figure \ref{H&rhozoom2}. That information is contained in figure \ref{Hvst}, which shows the way the Hubble rate changes as a function of cosmic time, for a solution to the effective equations (\ref{veq}) with $\phi(t_B)=1.2 m_{P\ell}$. Notice that while  figure \ref{H&rhozoom2} applies to all solutions, figure  \ref{Hvst} does not. However, the difference between  solutions is appreciable  only well after the bounce. Near the bounce, and during  the pre-bounce phase, quantum gravity effects dominate the spacetime dynamics, and  all solutions behave in the same way, regardless also of the form of the potential $V(\phi)$. This is to say, the fact that the de Sitter like contraction ends just a few Planck-times before the universe bounces is a universal feature of the Dapor-Liegener model.\footnote{The time at which the de Sitter phase ends, the value of the emergent cosmological constant, and the curvature at the bounce depend, however, on the value of the basic constants of the theory, namely $\gamma$ and $\ell_0$.}

\bfig\begin{center}
 \ig[width=0.7\textwidth]{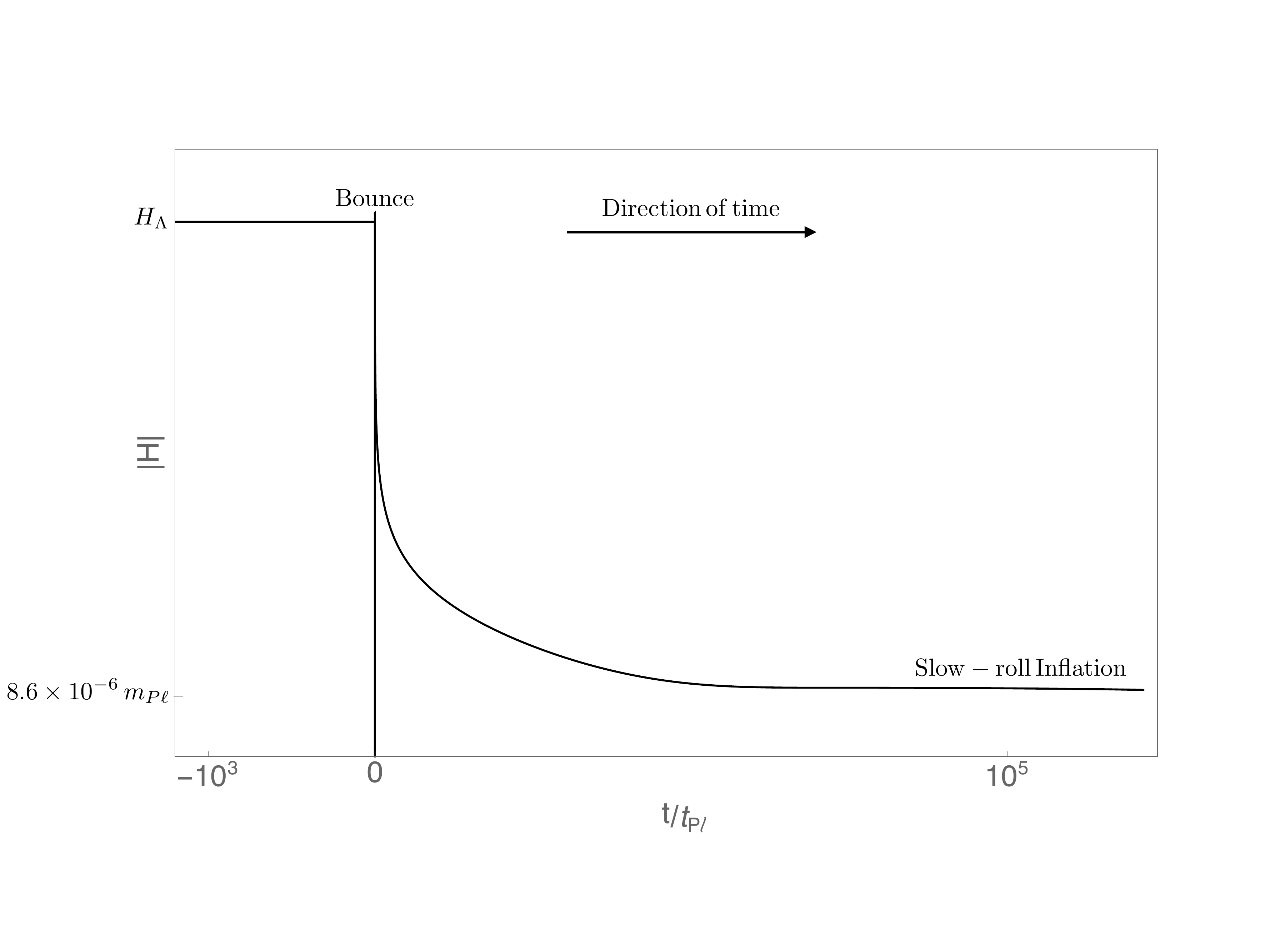}
 \ig[width=0.7\textwidth]{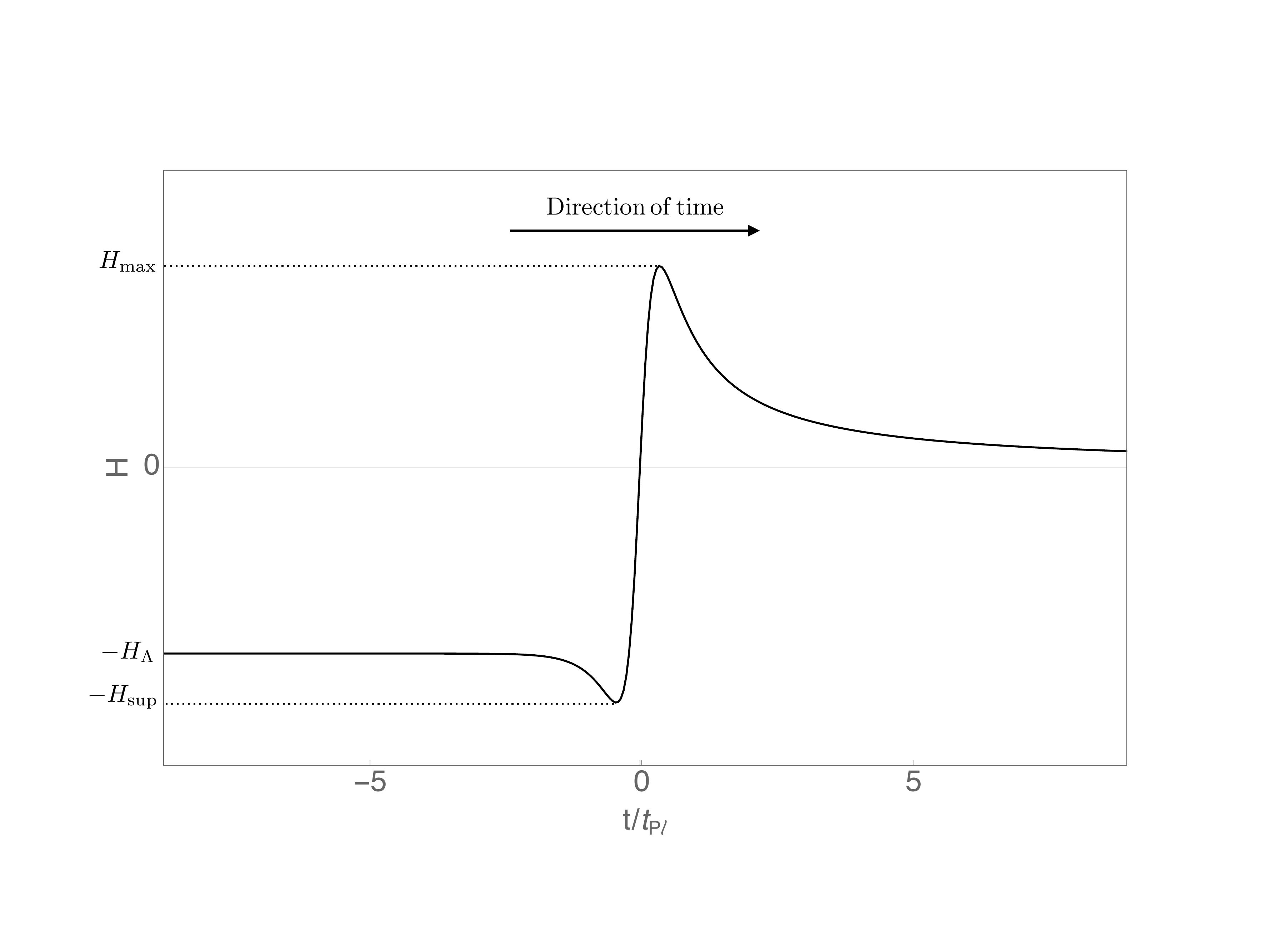}
\caption{Hubble $H$ rate versus cosmic time for the region of figure \ref{H&rhozoom2} corresponding to Branch 1, for a solution to the effective equations (\ref{veq}) with $\phi(t_B)=1.2 m_{P\ell}$. In contrast to figures \ref{H&rho} and \ref{H&rhozoom2}, in this plot time runs from left to right.  The upper panel shows the absolute value of $H$ for a  long time interval before and after the bounce. Before the bounce, $H$ becomes constant and of Planckian size. Well after the bounce, $H$ decreases until the onset of the inflationary era, during which it becomes almost constant, but with a value well below the Planck scale.\\ 
The bottom panels shows a zoom around the time of the bounce, and it makes evident how fast  $H$ reaches its asymptotic constant value before the bounce.}
\label{Hvst}\end{center}
\efig
 
\item{\bf Expanding phase} 

To the future of the bounce the evolution is very similar to standard LQC. Namely, both matter energy density and curvature decrease and vanish in the far future.  This can be seen from equations (\ref{rhoh}) and (\ref{Hh}), which show that $\ell_0h\to 0$, which corresponds to $t\to \infty$,  produces $\rho\to 0$ and $H\to 0$. Therefore, contrary to the pre-bounce phase, {\em there is no emergent cosmological constant after the bounce}. Immediately after the bounce, the Hubble rate grows from zero to a local maximum in around two Planck-times. After that, $H$ decreases, and quantum gravity effects quickly become subdominant once energy density and Ricci scalar are smaller than the Planck scale. Around one thousand Planck-times after the bounce,  general relativity becomes an excellent approximation. To the future, $\rho \to 0$ as $t\to \infty$, i.e., it is needed an infinite amount of cosmic time to reach $\rho=0$. This is equivalent to say that the regions $h>0$ and $h<0$ in figure \ref{H&rho} are {\em not connected by dynamics}; they correspond to different solutions of the equations (\ref{veq}). 

The scalar field  $\phi$ potential energy $V(\phi)$, plays a subdominant role on the  geometry at or before the bounce; quantum gravity effects dominate. This is no longer true  well after the bounce, when the evolution of the spacetime geometry becomes dominated by $V(\phi)$. If this potential has the appropriate form, it will eventually bring the Universe to a phase of slow-roll inflation \cite{Ashtekar:2011rm,Corichi:2010zp,Bonga:2015kaa}. 

\item{\bf $h<0$ region} 

This region  describes a different solution of the equations (\ref{veq}), which can be thought as the time reversal  of the solution described by the region $h>0$. The pre-bounce phase is almost identical to the one described by standard LQC. Namely, quantum gravity effects quickly become negligible to the past of the bounce, and general relativity becomes an excellent approximation for the dynamics. The remote past therefore describes a classical universe in which both the energy density and the Hubble rate vanish when $t\to - \infty$. 

After the bounce, however, one finds that an effective cosmological constant emerges. The Universe does not become classical in the future. Rather, the Hubble rate and the Ricci curvature quickly approaches a constant value $H_{\Lambda}\equiv\frac{1}{\ell_0}\frac{\gamma}{1+\gamma^2}$ and $R_{\Lambda}=12H_{\Lambda}^2=2.08 \ell_{P\ell}^{-2}$, and the expansion remains dominated by quantum gravity effects until $t\to \infty$. Matter and radiation will always be subdominant in this expanding branch, all the way up to $t\to \infty$, quickly diluting as the universe expands exponentially fast. Therefore, in such universe there would not be structure formation. Obviously, such a solution does not describe the cosmos we inhabit.

\end{enumerate}

To summarize, the solutions of equations (\ref{veq}) can be divided in two groups or branches. The first branch is compatible with our present universe, but has a non-singular `beginning'. The big bang singularity of general relativity is replaced by a quantum region in which the Hubble rate and the spacetime curvature remain constant and of Planck size all the way to $t\to -\infty$. Our classical universe emerges from a Planckian phase of constant curvature. 
The second branch of solutions is the `time reversal' of the first branch. \\

In the previous discussion, and in the computations showed in the next section, we will use the values of $\gamma$ and $\ell_0$ that are more commonly used in the literature. However, one could  also adopt a purely phenomenological viewpoint and compute the predictions of the model for different values of these parameters. This will change the value of both the emergent cosmological constant $\Lambda=3 H_{\Lambda}^2$ and the  matter energy density at bounce, $\rho_B$. Among all possible ways of changing $\gamma$ and $\ell_0$, there are two of particular  interest. 

\begin{enumerate}

\item Modifications of $\rho_B$ that leave $H_{\Lambda}$ unchanged. 

By modifying  $\gamma$ and $\ell_0$ while keeping the combination $\gamma/[\ell_0(1+\gamma^2)]$ unchanged, we obtain a different value of the energy density at the bounce while the  emergent cosmological constant $H_{\Lambda}$  remains unchanged.

\item Modifications of $H_{\Lambda}$ that leave $\rho_B$ unchanged. 

By modifying  $\gamma$ and $\ell_0$  keeping the combination $\ell_0(1+\gamma^2)$ unchanged, the energy density at the bounce  remains invariant, while the value of the cosmological constant $H_{\Lambda}$ changes.  In this case, for $\gamma\ll 1$, one obtains $H_{\Lambda}\propto \gamma$. Therefore, by choosing $\gamma$ small enough one could make $H_{\Lambda}$ to agree with the present  expansion rate, and hence account for the observed cosmological constant.  This would require $\gamma\approx 10^{-61}$. This is an obvious unnatural value for this parameter, so unnatural that would make quantum gravity effects to appear at extremely low curvatures.

\end{enumerate}

Furthermore, it is interesting to notice the existence a discontinuity in the value of the emergent cosmological constant $\Lambda=3\, H_{\Lambda}^2=3\,  \left(\frac{1}{\ell_0}\frac{\gamma}{1+\gamma^2}\right)^2$ in the limit $\ell_0 \to 0$. While $\Lambda$ grows when $\ell_0$ decreases, we see from equations (\ref{veq}) that if $\ell_0=0$ one recovers exactly general relativity {\em without} cosmological constant.\\

To finish,  it will important in the next section to keep in mind that solutions of the effective equations (\ref{veq}) are characterized by a single number---in addition to the fundamental parameters $\gamma$ and $\ell_0$ that appear in the Hamiltonian---which we choose to be the value of the scalar field at the time of the bounce $ \phi(t_B)$. To  understand why we only need one number to characterize a solution, although the phase space is four-dimensional, note first that in a spatially flat FLRW spacetime the scale factor $a$, or equivalently the ``volume" $v$, can be  re-scaled  without altering the physics. We choose $v=1$ at the bounce. On the other hand, at the bounce $\dot{{v}}=0$ {\it in all solutions}. Finally, because the energy density equals $\rho_{\rm sup}$ at the bounce in every solution, $ \phi(t_B)$ determines $\dot { \phi}(t_B)$. Therefore, from the apparently four initial data required to solve the system (\ref{veq}), the value of $ \phi(t_B)$ suffices to uniquely characterize a solution. Physically, $ \phi(t_B)$ dictates the amount of expansion that the universe accumulates after the bounce, which   grows very fast when $\phi(t_B)$ grows. This will make us interested in values of $\phi(t_B)$ that are not too much larger than 1, in Planck units, because otherwise the accumulated expansion  will be so large that all effects produced by the bounce will  be red-shifted out of our observable universe.

\section{The power spectrum of scalar perturbations}

We begin, in subsection \ref{initialstate}, by discussing the appropriate initial state for perturbations in the model under consideration. Then, we compute the power spectrum at the end of inflation in subsection \ref{PS}. We finish, in subsection \ref{comparison}, by comparing the results with the ones obtained from standard LQC, and by explaining  the origin of the differences.

To describe and evolve scalar cosmological perturbations we will use the so-called dressed effective metric approach, developed in \cite{Ashtekar:2009mb,Agullo:2012sh,Agullo:2012fc,Agullo:2013ai}. Scalar perturbations are described by a gauge invariant variable $\mathcal{Q}$. This variable is related to the familiar comoving curvature perturbations $\mathcal{R}$ by $\mathcal{Q}=\frac{z}{a}\mathcal{R}$, where $z=\dot \phi/H$.  In the dressed effective metric approach,  scalar perturbations are treated as Fock-quantum fields, and they obey the Klein-Gordon equation $[\Box-\u(\eta)]\mathcal Q=0$, where the Box operator $\Box$ is the D'Alembertian associated to the dressed effective metric defined e.g.\  in \cite{Agullo:2013ai}. In this paper, this metric is taken as the FLRW line element obtained from the effective equations (\ref{veq}). On the other hand,  the  time-dependent  potential $\u$ is given by $\u=a^2 [V(\phi)\,r - 2V_\phi(\phi)\sqrt{r} + V_{\phi\phi}(\phi)]$, where $r=3\dot \phi^2\,\frac{8\pi G}{\rho}$,  and $V_{\phi}(\phi)\equiv \dd V(\phi)/\dd\phi$ (see \cite{Agullo:2013ai} for additional details).\footnote{As discussed in  \cite{Agullo:2017eyh}, some quantization ambiguities appear in the definition of  $\u$. However, since these ambiguities produce effects in observable quantities that are  smaller than observational error bars, we do not discuss them here.}

In practice, the computation of the power spectrum follows the same steps as in semiclassical cosmology, with the important difference that the underlying metric does not satisfies Einstein's equations, but it is a solution of the equations (\ref{veq}). 

\subsection{Initial state for perturbations \label{initialstate}}

In contrast to the standard inflationary scenario, the model under consideration is free of any spacetime singularity, and the spacetime geometry is complete to the past. It is then natural to specify the initial state for cosmological perturbations in the asymptotic past. In that region, the Universe contracts exponentially fast in cosmic time, $a(t)=e^{-H_{\Lambda} \, t}$. This means that both the Hubble radius and the radius of curvature remain constant. On the contrary, wavelengths of Fourier modes grow exponentially fast when we propagate them back in time. Hence, all the modes we can observe in the CMB start their evolution when their wave wavelengths are much larger than the Hubble or curvature radius. This is very different from the situation in standard inflation, where  modes of interest for the CMB are deeply ``inside" the Hubble radius at the  onset of inflation. Recall that it is precisely the fact that  modes are well inside the Hubble radius at the onset of the slow-roll era that allows one to use adiabatic arguments to 
fix the initial state of perturbations in standard inflation. In the model under consideration, in sharp contrast, modes are not adiabatic at all in the past. Then, how can we specify the initial state? We will take advantage of the fact that  the  effective metric becomes {\em highly symmetric} in the past. Namely, it quickly approaches the FLRW-de Sitter metric soon before the bounce. %
It is well-known that the  isometries of FLRW-de Sitter suffice to single out a preferred vacuum state; this is the so-called Bunch-Davies vacuum \cite{Bunch:1978yq}.\footnote{The name ``Bunch-Davies'' vacuum is many times used in cosmology in a loose sense to refer to an adiabatic vacuum. However, the Bunch-Davies state was originally defined in a sharper way, namely it is the only vacuum state that is simultaneously invariant under all the isometries of the FLRW-de Sitter geometry and that is adiabatic (of infinite adiabatic order). These two conditions are enough to uniquely specify the state in the entire spacetime manifold, an not only its short distance behavior, as rather suggested by the terminology used in some cosmology literature.} This is the state we choose in the past. Therefore, the model under consideration contains elements to select initial conditions for perturbations in an unambiguous and elegant way.
Note that the spacetime metric before the bounce rapidly approaches {\em exact} FLRW-de Sitter. This contrasts with inflationary spacetimes when one only has a quasi-FLRW-de Sitter spacetime, where the deviation is characterized by the slow-roll parameters. 

To write the form of the Bunch-Davies vacuum, we use standard techniques of quantum field theory in curved spacetimes. Namely, we expand the  field operator $\Q(\vec{x})$ in Fourier modes
\be \label{RepQ}
        \Q(x) = \int\frac{d^3k}{(2\pi)^3}\,\Q_{\vec{k}}(\eta)\,e^{i{\vec{k}}\cdot{\vec{x}}} 
                  = \int\frac{d^3k}{(2\pi)^3}\,\Big[\hat A_{\vec{k}}\,q_k(\eta) + \hat A^\dagger_{-{\vec{k}}}\,q^*_k(\eta)\Big]e^{i{\vec{k}}\cdot{\vec{x}}} \,,  \ee
where the functions $q_k(\eta)$, labeled by $k\equiv |{\vec{k}}|$, are a complete set of solutions to the equation 
\be \label{modeqn} [\Box-\u(\eta)]\mathcal Q=0 \hspace{0.5cm} \longrightarrow\hspace{0.5cm} 
 q_k^{\prime\prime} + 2\frac{a '}{a} q_k^\prime + (k^2 +
a^2 \, \u)\,  q_k =0\, ,
\ee
 and, furthermore, they are chosen to be normalized by
\be q_k(\eta) q_k'^*(\eta)-q^*_k(\eta) q'_k(\eta) = \frac{i}{a(\eta)^2} \, , \ee
at any time. In the previous equations $\eta$ indicates conformal time, and a prime refers to a derivative with respect to it.   $\hat A_{\vec{k}} $ and $\hat A^\dagger_{\vec{k}}$ are creation and annihilation operators, satisfying  $[\hat A_{\vec{k}}, \hat A_{{\vec{k}}'}] = [\hat A^\dagger_{\vec{k}}, \hat A^\dagger_{{\vec{k}}'}] = 0$, $[\hat A_{\vec{k}} ,\hat A^\dagger_{{\vec{k}}'}] = \hbar \, (2\pi)^3 \delta({\vec{k}}-{\vec{k}}')$, and the vacuum is the state annihilated by all $\hat A_{\vec{k}} $. Equation (\ref{RepQ}) can be thought as the expansion of the field operator $\Q(x) $ in the basis $\{q_k(\eta)\}$, and from this  it is obvious that the operators $\hat A_{\vec{k}} $ and $\hat A^\dagger_{\vec{k}}$, and hence the definition of vacuum,  depend on the choice of basis  $\{q_k(\eta)\}$. The Bunch-Davies vacuum in a FLRW-de Sitter spacetime is obtained by choosing 
\be \label{BD}q^{\rm BD}_k(\eta)=H_{\Lambda}\, \frac{1}{\sqrt{2k^3}}(1+i\, k\eta)\, e^{-i\, k \eta}\, ,\ee
where the relation between the scale factor and conformal time in the FLRW-de Sitter epoch is $a(\eta) =\left |\frac{1}{\eta \, H_{\Lambda}}\right|$.

\subsection{Computation of  the primordial Power spectrum\label{PS}}

The primordial power spectrum of comoving curvature perturbations is given by
 \be \label{PS}
         P_{\mathcal{R}}(k) = \bigg(\frac{H(\eta)}{\dot\phi(\eta)}\bigg)^2 \hbar \frac{k^3}{2\pi^2}|q_k(\eta)|^2 \,.
    \ee
evaluated at the end of inflation. Then, in order to compute  $ P_{\mathcal{R}}(k)$ we first need to find the solutions $q_k(\eta)$ to  (\ref{modeqn}) associated with  the initial data given in $(\ref{BD})$ at a time well before the bounce, and substitute them in (\ref{PS}). These perturbations  propagate,  starting from the contracting phase of the universe, across the bounce and the post-bounce era, including the inflationary phase. The main difference with standard inflation is that the evolution from the remote past to the onset of inflation will modify the form of the modes, in such a way that they will reach the onset of slow-roll inflation in a quantum state that differs from the slow-roll Bunch-Davies vacuum. Therefore, the predictions for the primordial power spectrum will be different from the standard slow-roll predictions, as a consequence of the pre-inflationary history of the perturbations.

We solve the evolution of perturbations numerically. In order to perform these calculations we need to:

\begin{enumerate} 

\item Specify a potential $V(\phi)$ for the scalar field. In this paper we use the simple monomial potential $V(\phi)=\frac{1}{2}m^2\phi^2$, with the value of the mass derived from the Planck normalization \cite{Ade:2015lrj}, $m=1.25\times10^{-6}\, m_{P\ell}$. Other choices for $V(\phi)$ (compatible with slow-roll inflation) will leave our main conclusions unchanged. This is because the effects we  describe in this paper  originate from the Planck era of the universe---the bounce and the pre-bounce phase---and in that regime the contribution of the potential $V(\phi)$ to the dynamics is completely negligible. See \cite{Bonga:2015xna,Bonga:2015kaa,Agullo:2017eyh} for computations in LQC using the Starobisky potential. 

\item Specify a solution $v(\eta)$, or equivalently $a(\eta)$,  to the equations (\ref{veq}) that describe the spacetime geometry.  As discussed earlier, these solutions are uniquely characterized by specifying the value of  $\phi$ at the time of bounce, $\phi(\eta_B)$. 
In the plots below we use $\phi(\eta_B)=1.2 \, m_{P\ell}$, that corresponds to having $N_B=73.8$ $e$-folds  of expansion between the bounce and the end of inflation.  This choice determines  the size  of the wavelength that observable Fourier modes had at the time of the bounce. 

\item Specify the quantum state of scalar perturbations. As mentioned earlier, we use the Bunch-Davies vacuum in the far past.  At the practical level, it is of course not possible to start the numerical evolution in the infinite past, $\eta=-\infty$. Note however that during the FLRW-de Sitter era before the bounce, (\ref{BD}) is an exact solution of (\ref{modeqn}). Therefore, one can use this  analytic  form  to evolve the modes from  $\eta=-\infty$ until a  time close to the bounce, and then compute the evolution numerically  from that time to the future. 

\end{enumerate}
 
\bfig\begin{center}
 \ig[width=0.7\textwidth]{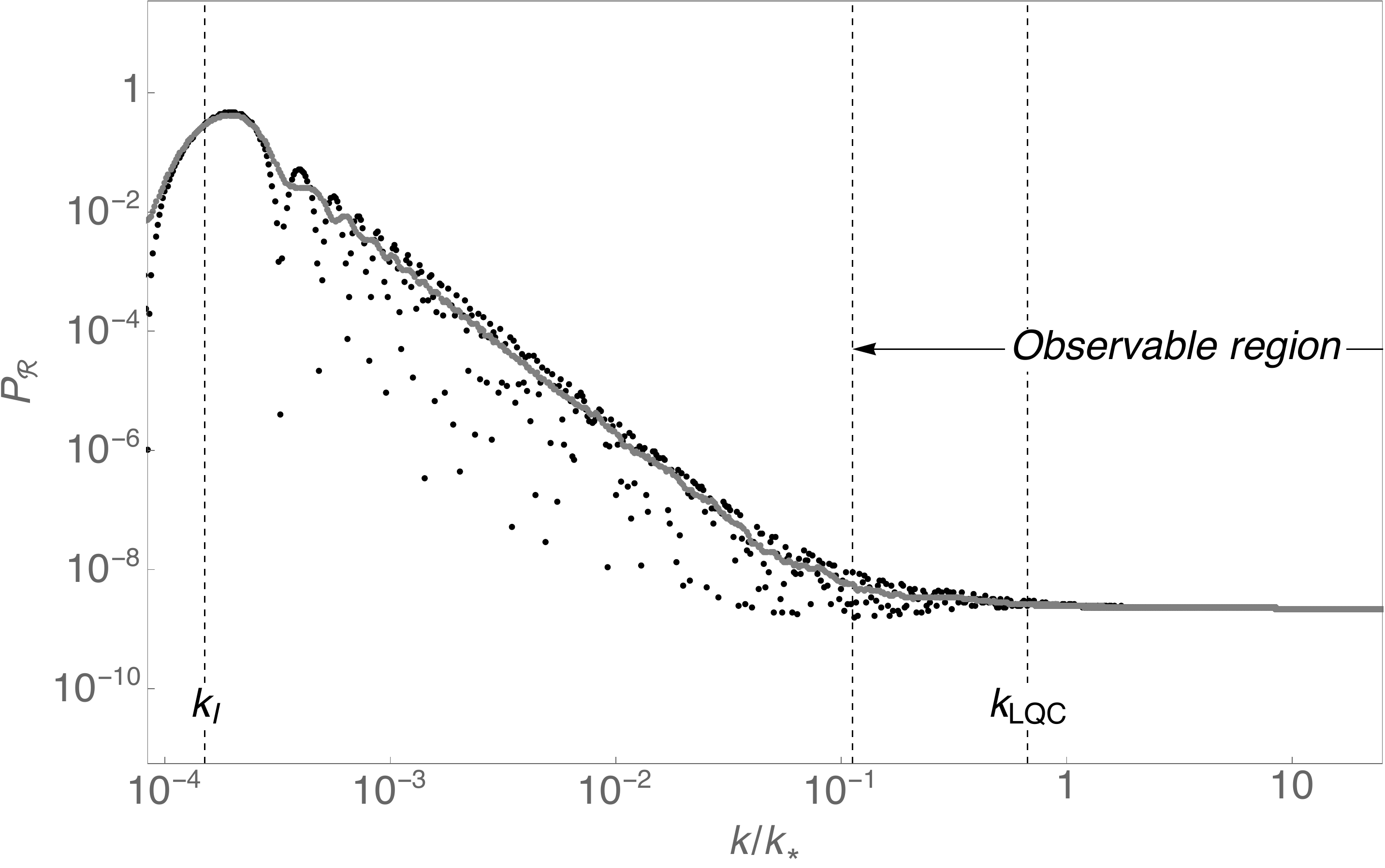}
\caption{Power spectrum of comoving curvature perturbations evaluated at the end of inflation. Black points show the value of the power spectrum  computed numerically for a set of wave-numbers $k$. The gray line is the average of the black points, obtained by binning the points in a window of Planckian size.
This plot has been obtained by using Bunch-Davies initial data for perturbations in the far past. The spacetime geometry  is a solution of the equations (\ref{veq}) with $\phi(t_B)=1.2 \, m_{P\ell}$. This corresponds to having 12.6 $e$-folds of expansion between the bounce and $t_{k_*}$---the time at which the reference mode $k_*$ that today has  physical magnitude $0.002\, {\rm Mpc}^{-1}$ exited the Hubble radius during inflation. The total number of $e$-folds from the bounce to the end of inflation is approximately $73.8$. For this spacetime, the  set of modes that we can  observe in the CMB is indicated in the figure, and ranges approximately from $k_*/9$ to $1000\, k_*$. The location of this observable window depends on the choice of $\phi(t_B)$, and it moves to left if we decrease $\phi(t_B)$. 
The figure also shows the location of the scale $k_{LQC}$ introduced by the bounce, and by the inflationary era, $k_{I}$.}
\label{NLQCPS}\end{center}
\efig

\begin{figure}\begin{center}
\includegraphics[width=0.7\textwidth]{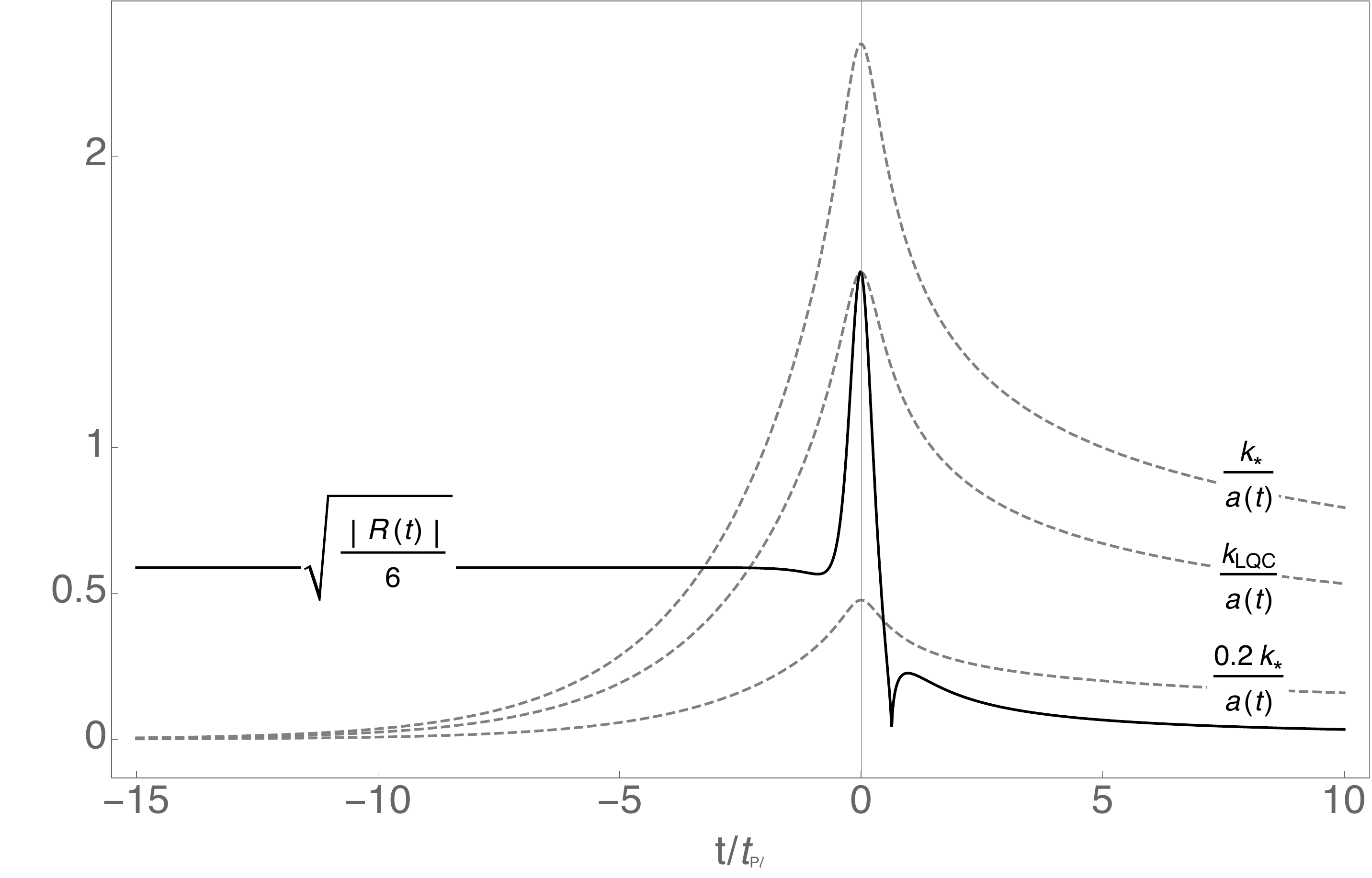}
\includegraphics[width=0.7\textwidth]{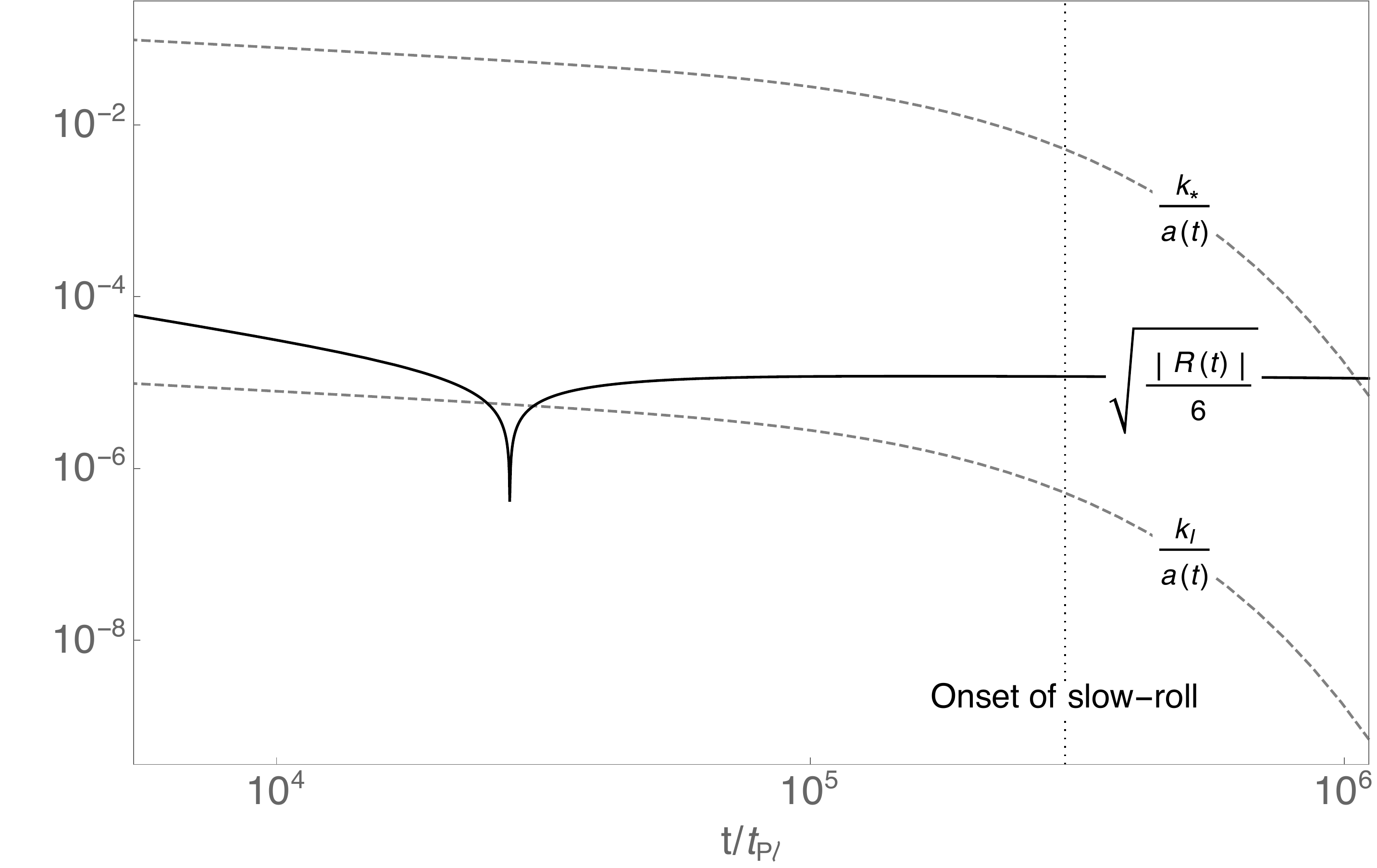}
\caption{Time evolution of several physical wave-numbers $k/a(t)$ and the scale $\sqrt{|R(t)|/6}$  ($R$ is the Ricci scalar) versus cosmic time $t$ in Planck units. \\
The upper panel shows the evolution around the time of the bounce. The most ultraviolet modes enter the curvature radius earlier. Very infrared modes (not shown) do not enter the curvature radius near the bounce. \\
The bottom panel shows the evolution near the onset of inflation. Observable modes exit the curvature radius during the slow-roll phase. Modes with $k\ll k_I$  remain outside the curvature radius during their entire evolution. }\label{Rsmodes}\end{center}
\end{figure}

The result for the primordial power spectrum is shown in figure \ref{NLQCPS}. The first lesson we learn is that the power spectrum looks qualitatively the same as in standard LQC---the next subsection  provides a quantitative comparison. %
Namely, the power spectrum is almost scale-invariant for the most ultraviolet wave-numbers (rightmost part of the plot), grows for  intermediate scales, and becomes very small for very infrared modes. The physical interpretation of this result is similar to the one in standard LQC (see e.g.\ \cite{Agullo:2015tca}),  and is obtained by paying attention to the way physical wave-numbers $k/a(\eta)$ evolve relative to the curvature of spacetime (see figure \ref{Rsmodes}).

\begin{itemize} 

\item{Ultraviolet modes.}
These are modes for which $k>k_{\rm LQC}$, where $k_{\rm LQC}$ is the scale defined by the bounce. More precisely\footnote{The factor $1/6$ is included in the definition of  $k_{\rm LQC}$ because it is  indeed the combination $R(\eta)/6$ what features in the equation of motion of perturbations. To see this explicitly, one needs to change variables to $v_k(\eta)\equiv a(\eta)\, q_k(\eta)$, for which the equation of motion takes the simple form $v_k''(\eta)+a^2(\frac{k^2}{a^2}+\u-\frac{R}{6})\, v_k(\eta)=0$. From this equation it is clear that the evolution of the modes $q_k(\eta)$ is dictated by the `competition' between the square of the physical wave-number $\frac{k}{a}$, the curvature scale $\sqrt{R/6}$, and the potential  $\u$. Furthermore, it turns out that the potential  $\u$ produces sub-leading contributions in the Planck era, relative to $\sqrt{R/6}$.} $k_{\rm LQC}\equiv a(\eta_B)\, \sqrt{R(\eta_B)/6}$, where $\eta_B$ indicates the time of the bounce. They  begin their evolution outside the curvature radius---i.e.\ $k/a(\eta)\ll  \sqrt{R(\eta)/6}$ in the past, and enter  during the contracting phase, when the universe is still in de FLRW-de Sitter phase. During this phase, the modes are exactly Bunch-Davies modes. But because  the Bunch-Davies vacuum is a state of infinite adiabatic order, they become  adiabatic modes when they are inside the curvature radius. Furthermore, these modes remain inside the curvature radius during the bounce, and  also during the pre-inflationary phase, to exit only during slow-roll inflation. Therefore, they reach the onset of inflation as adiabatic modes, and these are indistinguishable from the slow-roll Bunch-Davies modes at that time. This implies that their power spectrum at the end of inflation looks  the same as the standard inflationary power spectrum, namely  $ P_{\mathcal{R}}(k)=\frac{\hbar 4\pi G}{\epsilon(\eta_k)}\left(\frac{H(\eta_k)}{2\pi}\right)^2$, where  $\epsilon(\eta_k)$ and $H(\eta_k)$ are the slow-roll parameter and Hubble rate evaluated at the time $\eta_k$ the mode $k$ exits the Hubble radius during inflation. That is to say, ultraviolet modes do not acquire any imprint from the pre-inflationary evolution.

\item{Very infrared modes.}
These are modes for which $k<k_{\rm I}$, where $k_{\rm I}$  is a scale defined by Inflation. More precisely $k_{\rm I}\equiv a(\eta_I)\, \sqrt{R(\eta_I)/6}$, where $\eta_I$ indicates the beginning of the inflationary era, when the slow-roll parameter $\epsilon$ becomes smaller than one. It turns out that modes with  $k<k_{\rm I}$ never experience  curvature radius-crossing, neither in the pre-bounce and pre-inflationary era, nor during inflation.

\item{Intermediate modes.}
These are modes for which $k_{\rm I}<k<k_{\rm LQC}$. These modes enter the curvature radius at a time close to the bounce, when the universe is not anymore in the FLRW-de Sitter phase. This curvature radius-crossing amplifies them, and consequently they reach the onset of inflation in an excited state. This explains the enhancement of their power spectrum relative  to the standard inflationary predictions. This enhancement is not scale invariant, because different modes cross the curvature radius near the bounce at different times, and the curvature radius  changes quickly  during that epoch. 

\end{itemize}

The relevant  question now is to determine what is the size of the  wave-numbers that we can directly observe in the CMB  {\em relative to the scales of the problem} $k_{\rm LQC}$ and $k_{\rm I}$. The observable window is approximately given by $k\in [k_*/9,1000\, k_*]$, where $k_*$ is a reference mode  that today has  physical magnitude $k_*/a({\rm today})=0.002\, Mpc^{-1}$. In order to determine  what is the relation between $k_*/a(t)$ and  $k_{\rm LQC}/a(t)$, we need to propagate  $k_*/a(t)$ back in time until the bounce (or equivalently, to propagate $k_{\rm LQC}/a(t)$ forward until today). The result would obviously depend on the amount of expansion accumulated after the bounce. But remember that this expansion is dictated by our choice of $\phi(t_B)$, the value of the background scalar field  at the bounce. For the value used in figure  \ref{NLQCPS}, $\phi(t_B)=1.2 \, m_{P\ell}$, we have  $[k_*/a(t_B)]/[k_{\rm LQC}/a(t_B)]\approx 1.5$, and the observable window begins near the interval  where  the power spectrum becomes scale invariant, in such a way that the enhancement of power would be visible only for the most infrared modes in the CMB. Smaller values of  $\phi(t_B)$ would make the observable window to shift to the left in figure \ref{NLQCPS}, and would predict an observable  power spectrum with large deviations from scale invariance for all scales in the CMB, in contradiction with observations. On the other hand, larger values of $\phi(t_B)$ would make the observable window to shift to the right, making the predictions indistinguishable from the standard inflationary predictions. This is the reason why we are interested in values of $\phi(t_B)$ close to $1.2 \, m_{P\ell}$, since it is for them that we can have some new effects in the CMB, but only for the most infrared scales, avoiding a  violation of the observed scale-invariance.

\subsection{Comparing with standard LQC\label{comparison}}

Figure \ref{NLQSvsSLQCPS} compares the power spectrum obtained in the Dapor-Liegener model with Bunch-Davies initial conditions in the far past, and the power spectrum obtained from standard LQC \cite{Agullo:2013ai}. The later is computed by using the ``Minkowski vacuum'' at the time of the bounce.\footnote{As discussed in \cite{Agullo:2015tca},  similar results for the power spectrum are obtained in standard LQC by using a fourth order adiabatic vacuum either at the bounce, or far into the past before the bounce. We use here Minkowski-like initial data at the bounce because, as we will shortly see, it makes the comparison easier, but the conclusions would remain unchanged if any of these other initial states were used in standard LQC.} (See footnote \ref{Mvacuum} for a definition of ``Minkowski vacuum''.) Both  spectra are obtained by using the same value of the curvature  and the scalar field at the bounce, because only in this way the comparison is meaningful.  Figure \ref{NLQSvsSLQCPS}  contains two main messages. 

\bfig\begin{center}
 \ig[width=0.7\textwidth]{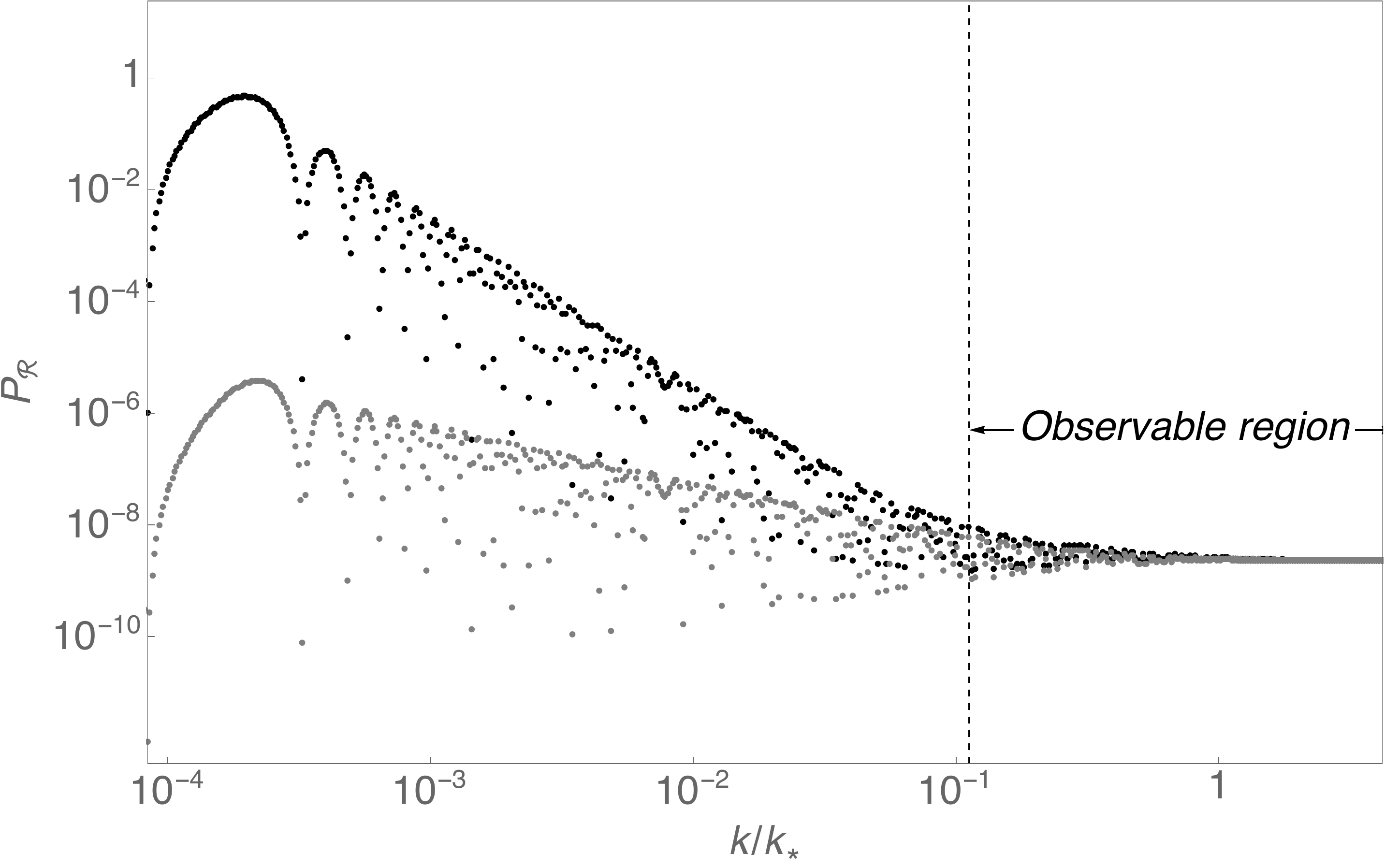}
 \ig[width=0.7\textwidth]{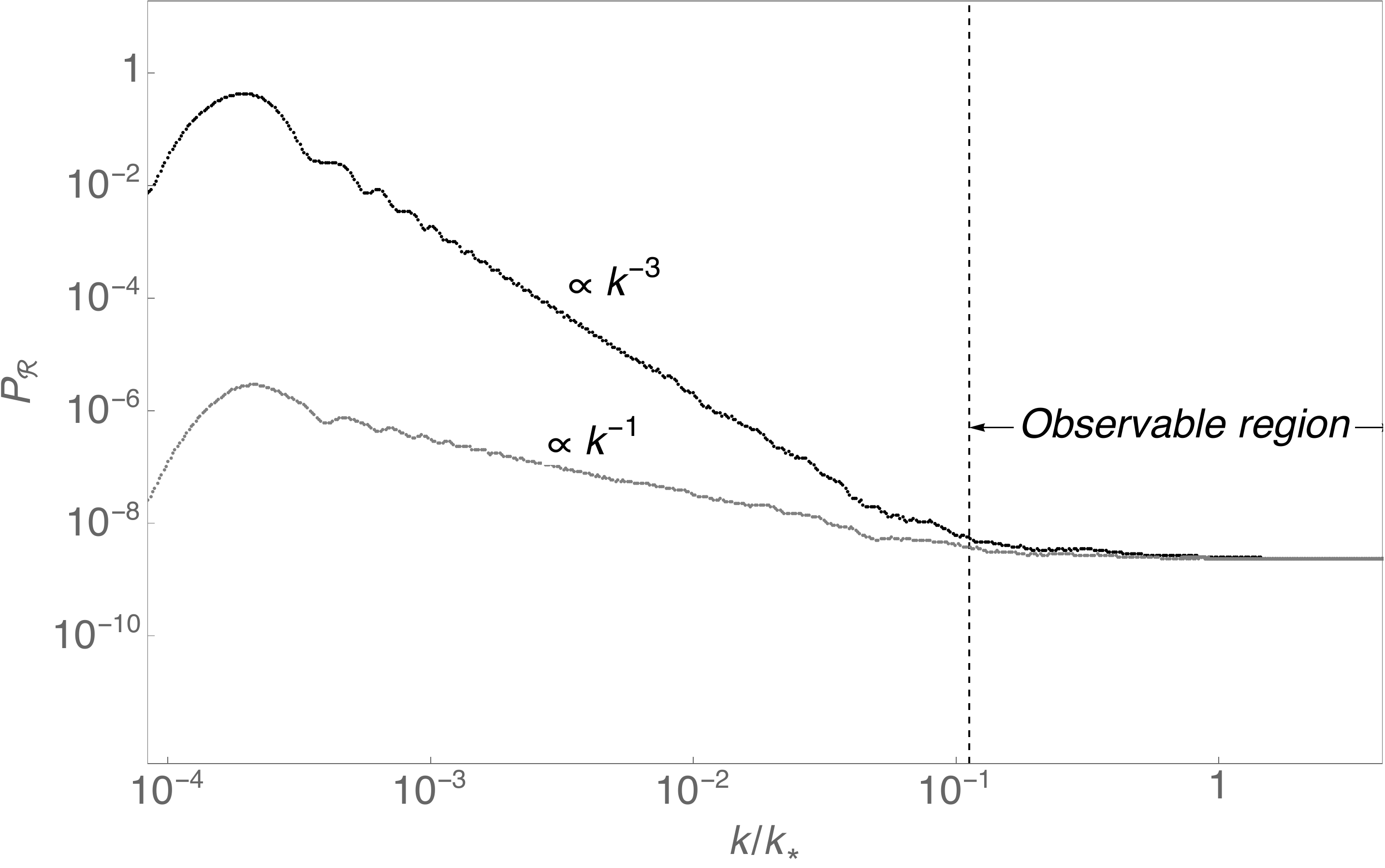}
\caption{Comparison of the power spectra obtained  in the Dapor-Liegener model (black) and in standard LQC (gray). Points in the upper panel shows the value of the two spectra  computed numerically for a set of wave-numbers $k$. The lower panel shows the average of the upper panel, obtained by binning the points in a window of Planckian size.  The two spectra have important differences for infrared sales, but agree for most of the wave-numbers that we can directly observe in the CMB.  
.}
\label{NLQSvsSLQCPS}\end{center}
\efig

\begin{enumerate}

\item  Both spectra are indistinguishable for ultraviolet modes. Differences between them only appear for modes well inside the intermediate region discussed above, $k_{\rm LQC}\gtrsim k\gtrsim k_{\rm I}$. For the parameters used in figure \ref{NLQSvsSLQCPS}, both spectra agree inside the observable window. But observable  differences would have appeared for the most infrared scales in the CMB if we had chosen a slightly smaller value of $\phi(t_B)$.

\item The main difference between the two spectra is a {\em different slope for intermediate scales}. The power spectrum in  the Dapor-Liegener mode grows faster $(\propto k^{-3})$ when we move towards infrared scales. As we explain further below, this is the distinctive characteristic  of the Dapor-Liegener model. 

\end{enumerate}

Because the differences between the two spectra are only for scales for which the power spectrum is strongly scale dependent, these differences can appear only in the most infrared scales of the CMB. But observational error bars are large for these scales. Hence,  it may be  challenging to distinguish both models by looking only at the CMB scalar power spectrum. However, the specific $P_{\mathcal{R}}\propto k^{-3}$ prediction of the Dapor-Liegener model for infrared modes may help on this task. 

Although we have only discussed so far the scalar power spectrum, our conclusions also apply for the tensor perturbations. Hence, observations of tensor modes will help to tests the predictions of this model. Furthermore, we expected that the non-Gaussianty will show even larger differences between the Dapor-Liegener model and standard LQC. However, the computation of  non-Gaussianity is  beyond the scope of this paper. 

In order to better  understand  the origin of the differences between both spectra, we  compute the Bogoliubov coefficients $(\alpha_k,\beta_k)$ that relate the modes $q_k(t)$ that result  from the Dapor-Liegener model with Bunch-Davies vacuum in the past, and the ones that are obtained in standard LQC from ``Minkowski-like" initial conditions at the bounce.\footnote{\label{Mvacuum}By Minkowski-like initial data we mean:  $q^{\rm M}_k(\eta_0)\equiv\frac{1}{a(\eta_0)\sqrt{2k}}$ and $q^{\rm M}_k(\eta_0)\equiv\frac{-i \, k}{a(\eta_0)\sqrt{2k}}$.  The name is motivated by the fact that this initial data would produce the familiar Minkowski vacuum in Minkowski spacetime.} Because  the spacetime geometry after the bounce is extremely similar in both models, one can obtain an excellent approximation for these coefficients by simply restricting ourselves to the Dapor-Liegener model, and computing the relation between the modes with Bunch-Davies initial data in the past and those with Minkowski initial data at the bounce. The result is shown in figure \ref{a+b2}, and reveals that these coefficient are non-trivial---i.e. $|\beta_k|\gtrsim 1$---only for sufficiently infrared wave-numbers, for which we find $|\beta_k|\propto k^{-2}$. For $k>0.2\, k_{\star}$ we find $|\beta_k|<0.1$. 

The form of the  coefficients $(\alpha_k,\beta_k)$ shown in figure \ref{a+b2} can be understood from the  the following  simple argument. 
Consider the Bogoliubov transformation between modes with Bunch-Davies form  (\ref{BD}) {\em very near the bounce}, and modes with Minkowski-like initial data at the same instant. These coefficients are not exactly the ones shown in figure \ref{a+b2}, since the later  refer to modes that have the Bunch-Davies form in the past, and the evolution until the bounce  slightly changes their form. But since the de Sitter phase ends just a few Planck seconds before the bounce, it is a good approximation to neglect this change. The advantage is that one can compute $(\alpha_k,\beta_k)$ analytically. We obtain $\beta_k=\frac{H\, a}{2 k}$ and $\alpha_k=\beta_k+i$. From this we see  explicitly that $|\beta_k|^2\propto k^{-2}$, and $|\beta_k|^2$ becomes smaller than 1 for sub-Hubble modes.  Although this analytical argument provides  only  an approximation, it helps to understand that the difference between the two spectra in figure \ref{NLQSvsSLQCPS} originates from the specific Bunch-Davies initial conditions used in the Dapor-Liegener model. Furthermore, since these initial conditions come from the emergent de Sitter phase, {\em the $\propto k^{-3}$ behavior of the infrared  part of the power spectrum is the fingerprint of the Dapor-Liegener model.} \\

\bfig\begin{center}
 \ig[width=0.7\textwidth]{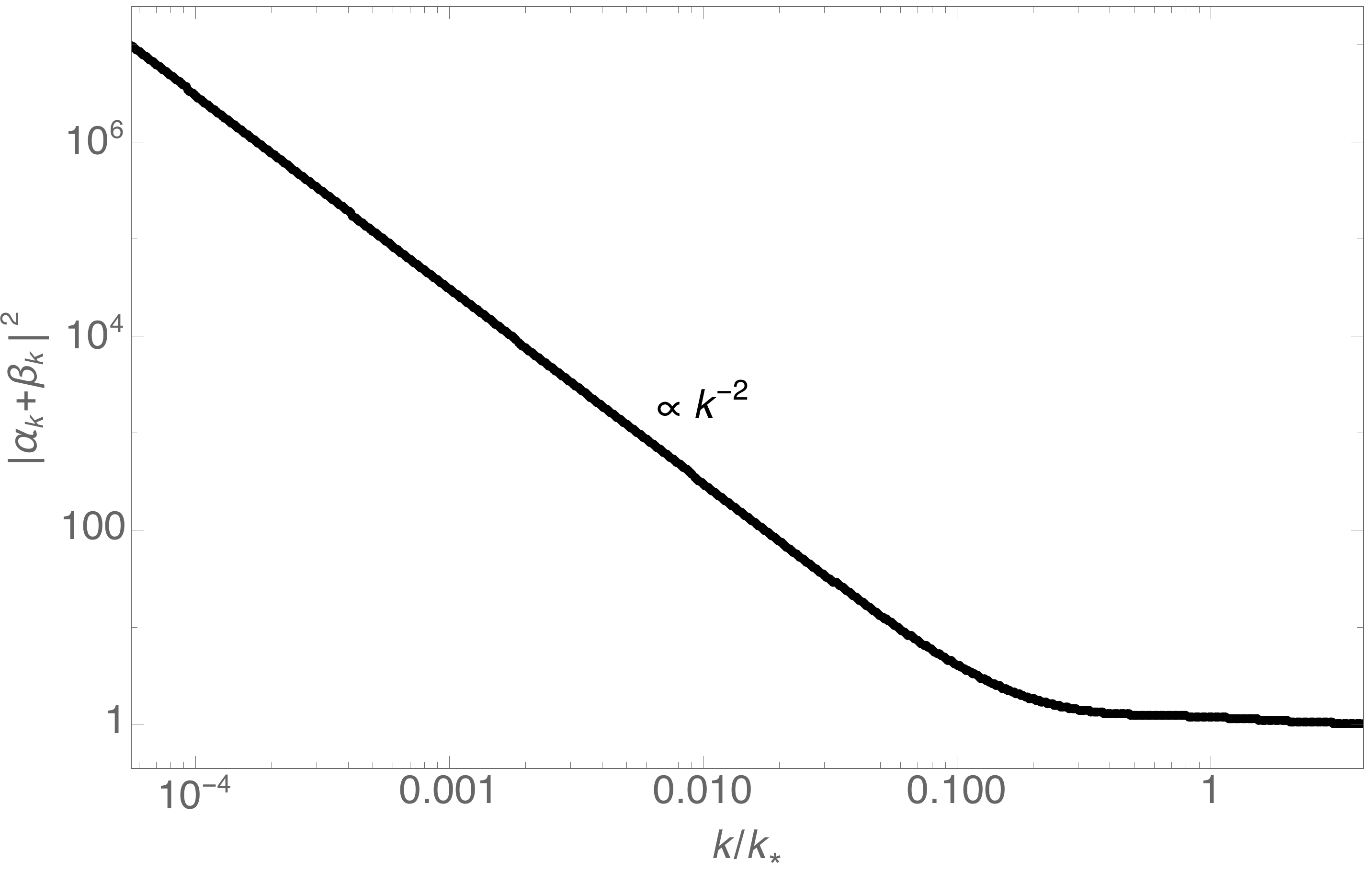}
 \ig[width=0.7\textwidth]{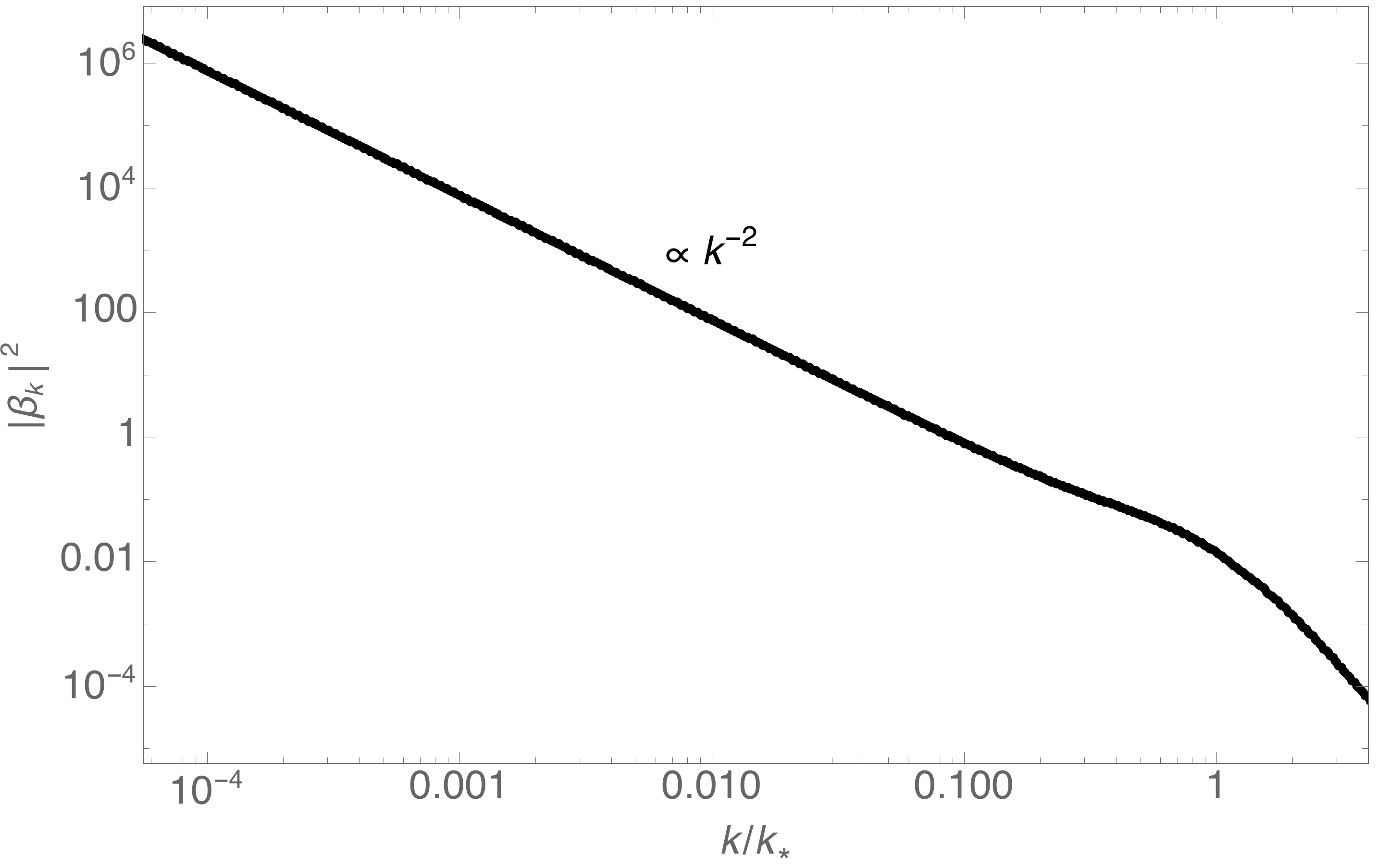}
\caption{Bogoliubov  coefficients $(\alpha_k,\beta_k)$ between the modes $q_k(t)$ used in the Dapor-Liegener model and in standard LQC. The upper panel shows $|\alpha_k+\beta_k|^2$, that turns out to be the same as the ratio between the power spectra of the two models. The bottom panel shows $|\beta_k|^2$, that physically represents the  relative number density of  quanta between the quantum states used in the two models.}
\label{a+b2}\end{center}
\efig

{\bf Remark.} The large values of the power spectrum  in the Dapor-Liegener model for very infrared scales  raises some concerns about the validity of the perturbative expansion for these scales. However, it is not obvious whether this large value is in fact  problematic, because they are extremely infrared modes, with very little energy on them. One would need to compute the next to leading order in perturbation theory in order to evaluate whether the perturbative expansion is jeopardized for these infrared scales. However, because these modes have wave-lengths that today are much larger than the radius of the observable universe,  they do not have  observable consequences, and can be absorbed in the background geometry. The form of their power spectrum and whether they can be described perturbatively, are irrelevant questions for both theory and observations.

\section{Conclussions}

Dapor and Liegener have recently  proposed a new Hamiltonian for the effective geometry of  loop quantum cosmology, by following  Thiemann's regularization of the Hamiltonian constraint of loop quantum gravity \cite{Dapor:2017rwv,Dapor:2017gdk}. These  authors,  in collaboration with  Assanioussi and Pawlowski, have further analyzed the details of this proposal, and put it in more solid grounds \cite{Assanioussi:2018hee}. The effective Hamiltonian (equation \ref{Heff}) is significantly more complicated than the one normally used in LQC. In particular, it contains a new term that is {\em quartic} in $\sin{(\ell_0\, h)}$, where $h$ is the conjugate variable of the volume $v$ (in the standard LQC, the effective Hamiltonian is only quadratic in $\sin{(\ell_0\, h)}$). These complications translate to new interesting physical features in the solutions; they make  our classical universe to emerge from a de Sitter-like phase of constant curvature. More precisely, the new proposal predicts that the big bang singularity is replaced by a quantum  quantum bounce. This is very similar to the predictions of standard LQC. But important differences appear in the pre-bounce universe. The new term in the Hamiltonian constraint makes a cosmological constant to {\em emerge} before the bounce. This is surprising, since there is no cosmological constant in the equations of motion;  it is rather a non-trivial combination of quantum gravity effects that acts as such.  Hence, the  pre-bounce, classical contracting universe of standard LQC is replaced by a region in which the spacetime curvature remains constant and of Planckian size, from $t\to -\infty$ until a just few Planck-times before the bounce, while the universe contracts exponentially fast in the forward-in time evolution. We find  remarkable the way in which the new complications in the Hamiltonian constraint manage to give rise to an emergent cosmological constant before the bounce, which furthermore {\em disappears after it}, making our classical universe to emerge at late times. 

The equations of motion also admit the time reversal as solutions. They correspond to classically contracting universe, which bounce and enter a de Sitter  expanding phase dominated by an emergent, cosmological constant of Planckian size until $t\to \infty$. These solution, however, do not describe our  classical universe.

In this paper we have analyzed the consequences that this new proposal has for the cosmic microwave background. Our findings can be summarized as follows. First of all, the contracting Sitter-like phase makes the past universe highly symmetric, and these symmetries  define a preferred  initial state for scalar perturbations  at $t\to -\infty$: the Bunch-Davies vacuum. Secondly, Fourier modes of perturbations start in the past with very  long wave-lengths, much longer than the radius of spacetime curvature. This contrasts with the situation in standard inflation where modes of interest start in the adiabatic regime. And third, the de Sitter phase of the early universe produces a very concrete feature in the power spectrum of scalar perturbations, namely it behaves as $P_{\mathcal{R}} \propto k^{-3}$  for  infrared modes (in contrast to  the $k^{-1}$ behavior obtained in standard LQC).  This prediction also extends to tensor modes, and constitute the distinctive feature  of the Dapor-Liegener model.

The results of this paper make manifest the way different quantization strategies within loop quantum cosmology can be brought to the realm of observations.  Observational differences are however restricted to the most infrared scales we could observed in the most optimistic scenario, or to unobservable super-Hubble scales otherwise. Furthermore, for  infrared scales in the CMB, observational error bars are large. In addition to tensor mode, predictions for non-Gaussianity would help in distinguishing the predictions of the Dapor-Liegener  from standard LQC, but their analysis is beyond the scope of this paper. But even with all these limitations in mind, we find important the fact that one can translate different proposals for the regularization of the Hamiltonian constraint into concrete features in the infrared sector of the CMB power spectrum. 
 
The analysis of this paper contains also  some limitations. First of all, as stated in the introduction, we have taken here a purely phenomenological approach, and not scrutinized in detail the limitations of the assumptions on which  the Dapor-Liegener proposal rests. Additional analysis, following \cite{Assanioussi:2018hee}, would help to firmly establish the mathematical and physical consistency of the model. Secondly, regarding perturbations, our analysis rests on the assumption that the back-reaction and the contribution of higher order perturbations to correlation function is subdominant. This justifies the use of dressed metric approach, in which one effectively works in a semiclassical theory.  Although these are reasonable assumptions, it would be desirable to explicitly check that this is in fact the case, in particular in the quantum de Sitter phase in which the spacetime curvature remains Planckian all the way to $t\to -\infty$. Such analysis, however, would require to include perturbations up to second order, as done for standard LQC in \cite{Agullo:2017eyh}. This will be the focus of future work.

\section*{Acknowledgments}
We thank A. Ashtekar, J. Olmedo, J. Pullin, S. Sahini,  and P. Singh for stimulating discussions. We specially thank A. Dapor and T. Pawloswski for discussions and initial collaboration on this project, and  A. Ashtekar for useful comments on the manuscript. This work has been supported by the NSF CAREER grant PHY-1552603,  and funds of the Hearne Institute for Theoretical Physics.

\bibliography{Refs}

\end{document}